\documentclass[runningheads]{llncs}
\usepackage{graphicx}
\usepackage{amsmath}
\usepackage{booktabs}
\usepackage{siunitx} 
\usepackage{multirow}
\usepackage{subfigure}
\usepackage{url}
\usepackage[misc]{ifsym}
\usepackage{bbding}

\begin{document}
\title{Computer-aided Tumor Diagnosis in Automated Breast Ultrasound using 3D Detection Network}

\author{Junxiong Yu\inst{1, 2}\thanks{Junxiong Yu and Chaoyu Chen contribute equally to this work.}, Chaoyu Chen\inst{1, 2\star}\and Xin Yang\inst{1, 2}\and Yi Wang\inst{1, 2}\and Dan Yan\inst{3} \and Jianxing Zhang\inst{3}\and Dong Ni\inst{1, 2} \Envelope}
\authorrunning{Yu et al.}
\titlerunning{Computer-aided Tumor Diagnosis in ABUS using 3D Detection Network}

% index{Yu, Junxiong}
% index{Chen, Chaoyu}
% index{Yang, Xin}
% index{Wang, Yi}
% index{Yan, Dan}
% index{Zhang, JianXing}
% index{Ni, Dong}

\institute{
\textsuperscript{$1$}National-Regional Key Technology Engineering Laboratory for Medical Ultrasound, School of Biomedical Engineering, Health Science Center, Shenzhen University, Shenzhen, China\\
\textsuperscript{$2$}Medical UltraSound Image Computing (MUSIC) Lab, Shenzhen University, China\\
\email{nidong@szu.edu.cn} \\
\textsuperscript{$3$}Guangdong Province Traditional Chinese Medical Hospital, GuangZhou, China\\ 
}
 \maketitle              % typeset the header of the contribution
\begin{abstract}
Automated breast ultrasound (ABUS) is a new and promising imaging modality for breast cancer detection and diagnosis, which could provide intuitive 3D information and coronal plane information with great diagnostic value.
However, manually screening and diagnosing tumors from ABUS images is very time-consuming and overlooks of abnormalities may happen.
In this study, we propose a novel two-stage 3D detection network for locating suspected lesion areas and further classifying lesions as benign or malignant tumors.
Specifically, we propose a 3D detection network rather than frequently-used segmentation network to locate lesions in ABUS images, thus our network can make full use of the spatial context information in ABUS images.
A novel similarity loss is designed to effectively distinguish lesions from background.
Then a classification network is employed to identify the located lesions as benign or malignant.
An IoU-balanced classification loss is adopted to improve the correlation between classification and localization task.
The efficacy of our network is verified from a collected
dataset of 418 patients with 145 benign tumors and 273 malignant tumors.
Experiments show our network attains a sensitivity of 97.66\% with 1.23 false positives (FPs), and has an area under the curve(AUC) value of 0.8720.
\end{abstract}

\section{Introduction}
For women all around the world, breast cancer is the most commonly diagnosed type of cancer.
Early detection through screening and advances in treatment have been shown significantly reduced the mortality rates.

Due to the advantages of non-invasive and convenient, ultrasound has become the most commonly used screening tool in the diagnosis of breast cancer, among which hand-held ultrasound (HHUS) is the most widely used.
However, HHUS has a high dependence on the diagnosticians and a low repeatability.
In contrast, automated breast ultrasound (ABUS) can make up for these shortcomings by providing more intuitive three-dimensional information and coronal plane information with great diagnostic value.

\begin{figure}[t]
    \centering
    \includegraphics[scale=0.55]{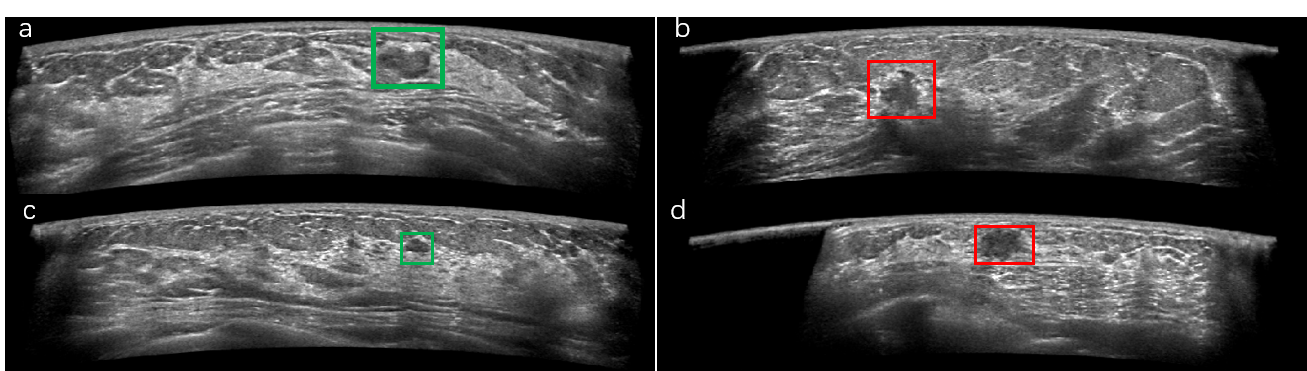}
    \caption{Example ABUS images, the area in the box is the lesion marked by the doctor, where a, c are benign lesions, and b, d are malignant tumors.}
    \label{fig:1}
\end{figure}

Although ABUS images have many advantages, they also inevitably increase the workload of doctors.
Generally a typical ABUS exam has at least three volumes to complete coverage of the entire unilateral breast.
Even for senior doctors, it is very time-consuming to manually screen tumors and overlook of abnormalities may happen. Therefore, the development of efficient and accurate computer-aided diagnosis is of great significance for reducing the workload of doctors, improving the tumor screening rate of ABUS images and promoting the early diagnosis of breast cancer. 

Nevertheless, developing computer-aided diagnosis (CAD) schemes for ABUS images remains challenging.
As shown in Fig.~\ref{fig:1}, 1) compared with other imaging modality, ultrasound imaging quality is relatively poor, thus making the boundary labeling difficult; 2) in most cases the proportion of lesion areas is less than 1\%, at the same time, the high similarity of benign and malignant lesions makes the classification task difficult; 3) the reconstructed ABUS images have approximately 800 frames, which requires huge computing resources.

In order to improve the efficiency of reviewing ABUS images, researchers have been developed many CAD systems.
Tan \textit{et al.}~\cite{tan2013computer} proposed an ensemble of neural network classifiers which obtains sensitivity of 64\% at 1 false positives (FPs) per image.
Lo \textit{et al.}~\cite{lo2014multi} proposed a CAD system based on watershed transform, achieving sensitivity of 100\%, 90\% and 80\% with FPs of 9.44, 5.42, and 3.33, respectively.
Moon \textit{et al.}~\cite{moon2014tumor} proposed a CAD system based on quantitative tissue clustering algorithm to identify tumors, achieving sensitivity of 89.19\% with 2.0 FPs per volume.
Wang \textit{et al.}~\cite{wang2019deeply} employed  convolutional neural networks (CNNs) with threshold loss for cancer detection, obtaining a sensitivity of 95.12\% with 0.84 FPs per volume.
Chiang \textit{et al.}~\cite{chiang2018tumor} applied 3D CNN and prioritized candidate aggregation, achieving sensitivities of 95\%, 90\%, 85\% and 80\% with 14.03, 6.92, 4.91, and 3.62 FPs per volume, respectively.
Moon \textit{et al.}~\cite{moon2020computer} proposed a 3D CNN with focal loss and ensemble learning, obtaining a sensitivity of 95.3\% with 6.0 FPs.
Wang \textit{et al.}~\cite{wang2020breast} proposed a CNN model which employs a multi-view strategy to classify breast lesions, obtaining AUC value of 0.95 with the sensitivity of 86.6\% and specificity of 87.6\%.

It can be found that most of the deep neural networks are based on U-Net~\cite{ronneberger2015u} architecture.
However, U-Net architecture consumes a lot of computing resource in the decode stage, which means only the small patch can be input into the network.
Therefore we propose a 3D detection network to make full use of the spatial context information in ABUS images.
In most traditional detection networks, regression and classification are two branches in parallel.
Many paper~\cite{pang2019libra,wu2019iou} have proved that enhancing the relationship between regression and classification will improve the performance of the network.
We propose to use IoU-balanced classification loss to make those anchors with high scores and good regression contributing more to the network.
To better distinguish the lesions from the background areas, we employ the similarity loss to increase intra-category correlation and inter-category discrimination.
After locating the lesions, we use a classification network to predict the class of lesions.

\begin{figure}[t]
    \centering
    \includegraphics[scale=0.142]{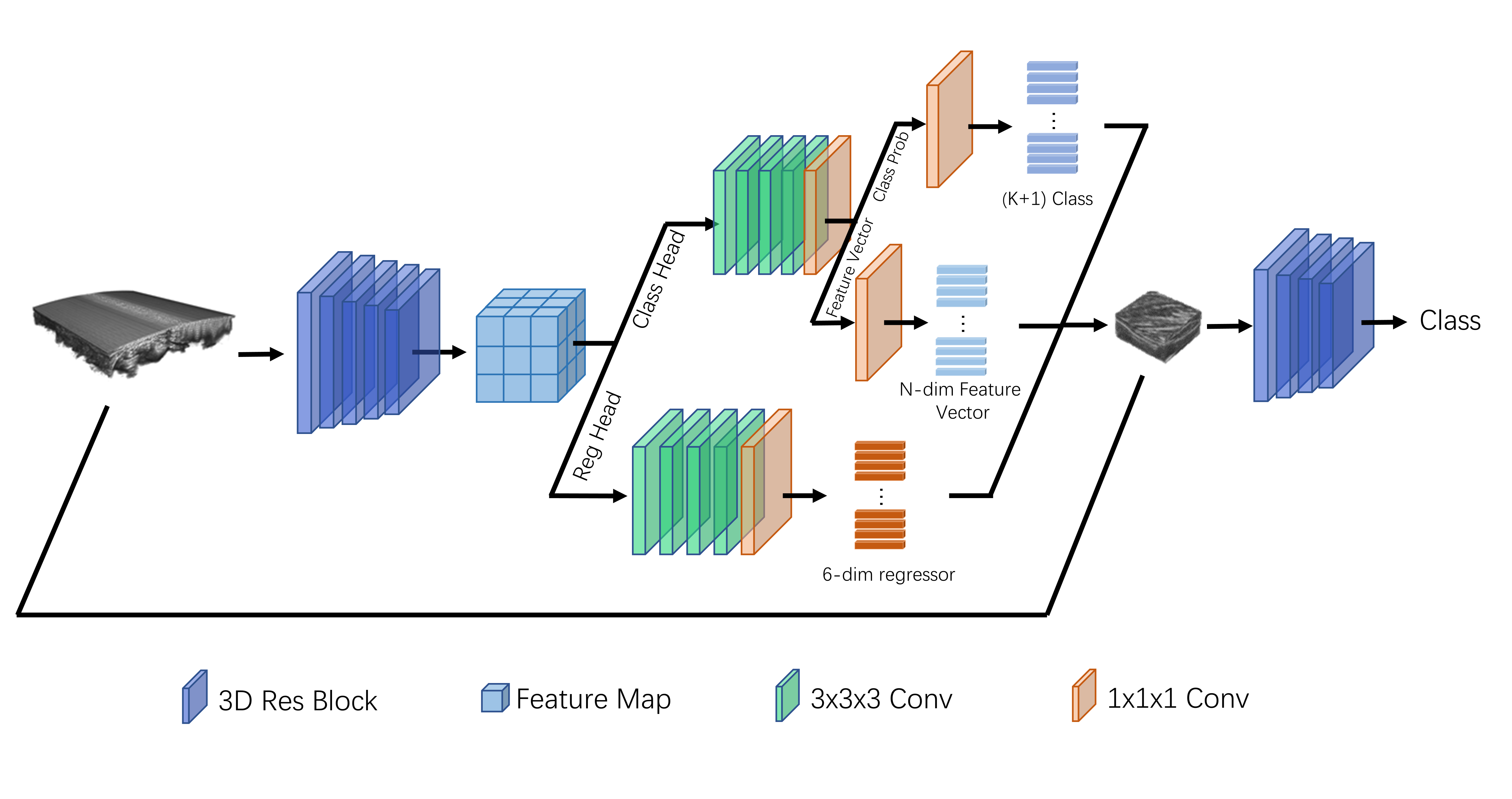}
    \caption{An overview of our proposed two-stage 3D detection network. A 3D region proposal network (RPN) is employed to locate suspected lesions, and then a classification network is used to identify benign or malignant tumors.}
    \label{fig:2}
\end{figure}

\section{Method}
In this section, we present the proposed two-stage 3D detection network.
Fig.~\ref{fig:2} provides an overview of our method, which leverages the backbone to extract the feature maps from the input ABUS images.
Then the feature maps are input into a 3D region proposal network (RPN)~\cite{ren2015faster} to locate suspected lesions, finally a classification network is employed to predict the class of these suspected areas.

\subsection{3D RPN}
To extract representative features, we use a 3D CNN consisting of 5 Res-blocks as the backbone.
The last feature map is input into a 3D RPN to locate suspected lesion areas.
After analyzing the size of lesions, we use 5 basic sizes (i.e., 8, 16, 28, 40 and 55) to generate 125 different anchors at each feature cell.
3D RPN is comprised of classification and regression branches, both of which contain four \begin{math}3\times 3\times 3\end{math} convolution layers.
At the head of regression branch, a \begin{math}1\times 1\times 1\end{math} convolution layer is employed to predict a set of six real-valued numbers representing bounding-box positions of the classes.
We use a \begin{math}1\times 1\times 1\end{math}  convolution layer to predict the probability of (K+1) classes of each anchor, meanwhile using a \begin{math}1\times 1\times 1\end{math} convolution layer to encode the corresponding area of each anchor to generate an n-dim (i.e., 32, 64, 128) feature vector.

\subsection{IoU-balanced Classification Loss}

In ABUS images, the lesion area often only accounts for a small proportion of the entire image, and is very similar with the background area.
Therefore, strengthening the classification weight of those anchors that regress well also helps to improve the performance of the network.
Thus, we propose to use IoU-balanced loss as the classification loss:
\begin{equation}
\label{eq.1}
L_{cls} = \sum_{i\in Pos}^{N}\omega_i(iou_i)*CE(p_i, \widehat{p_i}) + \sum_{i\in Neg}^{M}CE(p_i, \widehat{p_i}),
\end{equation}
\begin{equation}
\label{eq.2}
\omega_i(iou_i) = iou_i^\eta*\frac{\sum_i^n CE(p_i, \widehat{p_i})}{\sum_i^n iou_i^\eta*CE(p_i, \widehat{p_i})},
\end{equation}
In equation \ref{eq.1}, $iou$ is the Intersection-over-Union between positive proposal and its corresponding ground truth, $CE$ means cross entropy loss where $p_i$ is the predicted probability vector and $\widehat{p_i}$ is the real distribution, $\omega_i(iou_i)$ is the IoU weight from Eq~(\ref{eq.2}). The parameter $\eta$ can regulate IoU-balanced classification loss to focus on samples with high IoU and suppress the ones with low IoU. When $\eta$ is assigned to 0, the IoU-balanced classification loss is equivalent to cross entropy classification loss. 

We calculate the IoU between the positive regression bounding boxes and their corresponding ground truth boxes as the weight coefficient, which acts on the classification loss. On the one hand, the relationship between the regression branch and classification branch is strengthened; on the other hand, compared with the traditional cross-entropy loss for positive samples, IoU-balanced classification loss will get a higher weight coefficient for those positive samples which get higher IoU, thus when the network is updated, those positive samples with good regressions are more inclined to obtain higher classification scores. At the same time, those samples with poor regressions would get smaller weight coefficients, which suppress the impact of those samples with high classifications scores but poor regressions on the network.

\subsection{Similarity Loss}
Because of the characteristics of wide intra-class differences and small inter-class differences in ABUS images, in order to better locate and classify the lesions, an essential goal of our model is to learn the common characteristics of the same category as much as possible and expand the differences between different categories. 

Inspired by~\cite{chen2020simple}, we propose a similarity loss.
After encoding the area corresponding to the anchor into N-dimensional feature vectors, we select several feature vectors that match predetermined conditions.
The specific condition is that the IoU between each anchor and its corresponding ground truth needs to be greater than a certain threshold (i.e., 0.3), and the IoU between anchors also need to be greater than a certain threshold (i.e., 0.2).
After selecting these feature vectors that meet above conditions, we use Eq~(\ref{eq.3}) to calculate the cosine similarity between $Z_i$ and $Z_j$. Specifically, we calculate the cosine similarity between these selected vectors as $Sim_{pos,pos}$. Then we randomly select the same number of negative sample feature vectors and calculate the similarity between the negative sample feature vectors and the positive vetcors as $Sim_{pos,neg}$.
Our goal is to maximize the similarity between positive samples and reduce the similarity between negative samples, thus our loss function is as follows:
\begin{equation}
\label{eq.3}
Sim_{i, j} = \frac{Z_i^T Z_j}{\Vert{Z_i}\Vert*\Vert{Z_j}\Vert},
\end{equation}

\begin{equation}
L_{sim} = \frac{2-\log{(e^{Sim_{pos, pos}}/{e^{Sim_{pos, neg}}}})}{4}.
\end{equation}

The total 3D RPN loss is then summarized as
\begin{equation}
L_{rpn} = L_{reg} + L_{cls} + \lambda*L_{sim},
\end{equation}
where the regression loss is smooth L1 loss, $\lambda$ (= 0.7 in our implementation) balances the importance between $L_{cls}$ and $L_{sim}$.

\subsection{Lesion Classification}
We observe that the classification of the lesion is still insufficient if only using the output of the classification branch of the 3D RPN network.
Therefore, in order to predict the lesion category more accurately, we input the predicted candidates into a trained classification network similar to backbone, to predict its possible category.
%Data augmentation which included rotate, flipand is applied to prevent overfitting since there were almost twice as many negative cases as positives.
We then weigh the score of the detection network and the classification network to ascertain the final category.

\section{Experimental Results}
\subsection{Materials and Implementation Details}
Our experimental data were acquired from Sun Yat-Sen University Cancer Center. Our institutional review board approved the consent process.
There are totally 145 benign patients and 273 malignant patients involved in this study.
Each patient was scanned about 6 to 12 volumes with the voxel resolution of 0.511mm, 0.082mm, and 0.200mm in the transverse, sagittal and coronal direction, respectively.
We randomly divided the dataset: 250 patients in the training set, 84 patients in the validation set, and 84 patients in the test set. In test set, 84 patients have 251 volumes with 257 lesions (144 maligant and 113 benign).
All lesions were manually annotated by an experienced clinician.
Since ABUS data itself has a large scale (i.e., \begin{math}800\times 200\times 800\end{math}) thus is limited by the size of a single GPU memory, we down-sampled the raw volume to \begin{math}\frac{1}{8}\end{math} of its original size 
(i.e., \begin{math}400\times 100\times 400\end{math}).

During the training phase, we firstly randomly cropped a volume of \begin{math}400\times 98\times 360\end{math} around the lesion and then randomly cropped a volume of \begin{math}320\times 96\times 320\end{math} into the network.
Such operation can ensure that the input image maintains a high resolution, and can contain as many lesion areas as possible.
We specified an anchor as positive if it had the highest IoU with the ground truth or its IoU with ground truth was above 0.2.
An anchor was considered as negative if its IoU with every ground truth was less than 0.1.
Other anchors would be ignored in this study.

During testing phase, we got 4 patches of size \begin{math}320\times 96\times 320\end{math} from ABUS volume through regular crop.
For each patch, we only output the three boxes with the highest scores after non maximum suppression (NMS), and then the relative coordinates of the output box were converted into absolute coordinate.
The oversized or undersized prediction bounding-box were removed through post-processing.
The final prediction was retained after the NMS operation, and the maximum IoU is calculated for the reserved bounding-box.

The evaluation metrics consist of {$\textrm{mIoU}$} (the mean IoU across all categories), {$\textrm{FPs}$} (the number of false positives in a single data which the IoU between the ground truth is 0), and sensitivity.

\begin{table}[t]
	\centering
	\caption{Quantitative evaluation of our proposed framework.}
	\renewcommand\tabcolsep{6.5pt}
	\begin{tabular}{c|ccc}
		\hline 
		Method&$\textrm{mIoU} (\%)$ &$\textrm{FPs}$&$\textrm{Sensitivity} (\%)$ \\ 
		\hline
		2D-U-Net  &\textbf{44.34} &3.18  & 85.59\\
		3D-U-Net  &41.77 &2.73 & 87.04\\
		RPN     &36.49 &1.22 & 94.55 \\
		RPN-IoU &37.25 &1.37&96.11\\ 
		RPN-IoU-Sim&41.47 &\textbf{1.24}& \textbf{97.66}\\
		\hline
	\end{tabular}
	\label{table1}
\end{table}

\subsection{Performance Evaluation}
\subsubsection{Detection Results} Table~\ref{table1} shows quantitative comparison between the proposed framework (RPN-IoU-Sim, ``IoU'' denotes IoU-balanced classification loss; ``Sim'' denotes similarity loss) and other methods.
Compared with the traditional segmentation algorithm 2D U-net and 3D U-net~\cite{cciccek20163d}, the proposed class-specific RPN method has achieved better results, with a hit rate of 94.55\% and mIoU of 36.49\%.
While enhancing the relationship between classification and regression, RPN-IoU improves the performance of the RPN network, with a sensitivity of 96.55\% and mIoU of 37.25\%.
After using IoU-balanced classification loss, our detection performance has obviously improved in sensitivity and mIoU.
Finally, by combining IoU-balanced classification loss and similarity loss, our RPN-IoU-Sim network achieves a sensitivity of 97.66\% and mIoU of 41.47\%. Experimental results show that the proposed 3D detection scheme can achieve superior performance when using both IoU-balanced classification loss and similarity loss.

\begin{figure}[t]
	\centering
	\includegraphics[scale=0.4]{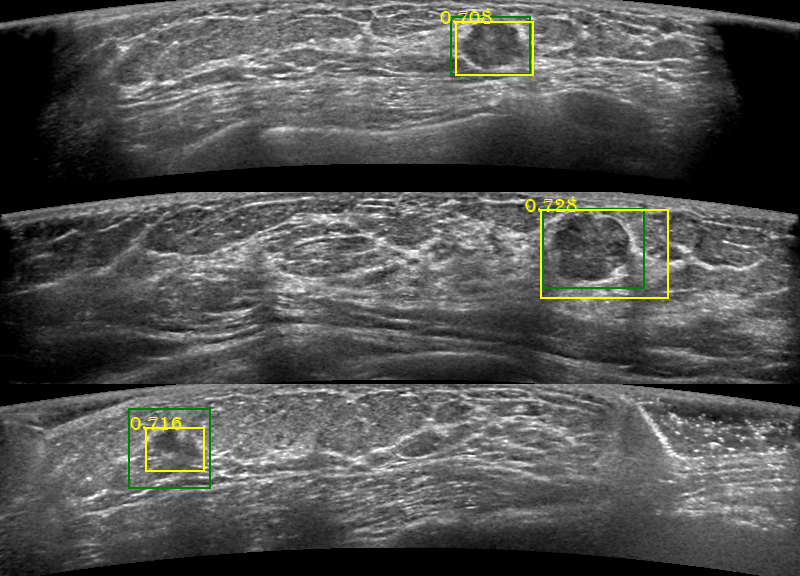}
	\caption{Example of lesion detection results. The areas in green boxes are the labels by doctors. The yellow boxes are model predicted with prediction probability values.}
	\label{fig:3}
\end{figure}

\begin{figure}[t]
	\centering
	\includegraphics[scale=0.4]{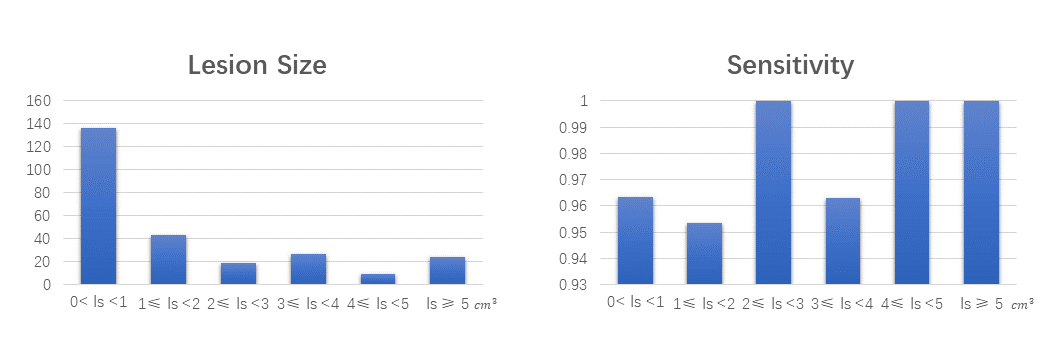}
	\caption{Left: the lesion size distribution of all lesions. Right: the detection sensitivities of different lesions. }
	\label{fig:4}
\end{figure}

Fig.~\ref{fig:3} shows three lesion detection results.
Fig.~\ref{fig:4} shows the sensitivity of our network to different sizes of lesions. For lesions smaller than 2 $cm^3$, our network achieved a sensitivity above 95\%; and when the lesion sizes was larger than 4 $cm^3$, the sensitivity is 100\%

\begin{table}[t]
	\centering
	\caption{Quantitative evaluation of our classification network.}
	\renewcommand\tabcolsep{6.5pt}
	\begin{tabular}{c|cccc}
		\hline
		Method &  Accuracy & Sensitivity & Specificity  & AUC\\
		\hline
		RPN & 0.7341 & 0.8241 & 0.6667 & 0.8154\\
		RPN-IoU & 0.8016 & 0.9259 & 0.7083 & 0.8628\\
		RPN-IoU-Sim &0.8016 & 0.9537 & 0.6875 & 0.8720\\
		\hline
	\end{tabular}
	\label{table2}
\end{table}

\subsubsection{Classification Results}
Table~\ref{table2} shows the quantitative comparison of classification results.
The RPN-IoU-Sim network outperformed the basic 3D RPN with respect to all evaluation metrics.

\section{Conclusion}
In this paper, we propose a 3D detection network for locating suspected lesions and classifying lesions as benign or malignant.
In the proposed network, we use a 3D detection network rather than frequently-used segmentation network to locate lesions in ABUS images.
By handling larger input patch, our network can make full use of the spatial context information in ABUS images.
Furthermore, IoU-balanced classification loss is employed to improve the sensitivity greatly by leveraging the correlation between classification and localization tasks.
Meanwhile, similarity loss is designed to effectively distinguish lesions from background.
Experimental results show our network obtains a sensitivity of 97.66\% with 1.23 FPs per ABUS volume and with an AUC value of 0.8720. 

\section*{Acknowledgements}
This work was supported in part by the National Key R\&D Program of China (No. 2019YFC0118300), 
in part by the National Natural Science Foundation of China under Grant 61701312, 
in part by the Guangdong Basic and Applied Basic Research Foundation (2019A1515010847), 
in part by the Medical Science and Technology Foundation of Guangdong Province (B2019046), 
in part by the Natural Science Foundation of SZU (No. 860-000002110129), 
and in part by the Shenzhen Peacock Plan (KQTD2016053112051497).


\bibliographystyle{splncs04}
\bibliography{paper2370}
%\printbibliography

\end{document}